# Evaporation and growth of crystals - propagation of step density compression waves at vicinal surfaces

Bogdan Ranguelov and Stoyan Stoyanov
Institute of Physical Chemistry, Bulgarian Academy of Sciences

rangelov@ipc.bas.bg; stoyanov@ipc.bas.bg

Abstract
   We studied the step dynamics during crystal sublimation and growth in the limit of fast surface diffusion and slow kinetics of atom attachment-detachment at the steps. For this limit we formulate a model free of the quasi-static approximation in the calculation of the adatom concentration on the terraces at the crystal surface. Such a model provides a relatively simple way to study the linear stability of a step train in a presence of step-step repulsion and an absence of destabilizing factors (as Schwoebel effect, surface electromigration etc.). The central result is that a critical velocity of the steps in the train exists which separates the stability and instability regimes. Instability occurs when the step velocity exceeds its critical value $V_{cr} = \dfrac{12 K \Omega A}{kTl^3}$ where $K$ is the step kinetic coefficient, $\Omega$ is the area of one atomic site at the surface, and the energy of step-step repulsion is $U = \dfrac{A}{l^2}$ where $l$ is the interstep distance. Integrating numerically the equations for the time evolution of the adatom concentrations on the terraces and the equations of step motion we obtained the step trajectories. When the step velocity exceeds its critical value the plot of these trajectories manifests clear space and time periodicity (step density compression waves propagate on the vicinal surface). This ordered motion of the steps is preceded by a relatively short transition period of disordered step dynamics.

1. Introduction
   Crystal sublimation takes place via detachment of atoms from the steps that exist at the vicinal surfaces. These atoms migrate on the terraces and eventually leave the crystal surface by desorption. Some of the adatoms can attach again to the elementary steps but the number of atoms reintegrated into the crystal is smaller than the number of atoms leaving the steps. As a result the steps move to the direction of the higher terraces. In general, the rate of the motion of a given step depends on the widths of the two neighbouring terraces. In the simplest case of perfectly regular step distribution all steps move at the same rate since all terraces have the same width. The step dynamics is extremely simple.

In reality, however, the steps are deviated from their perfectly regular positions. Intuitively, one expects these deviations to decrease and eventually vanish during the crystal sublimation because of the repulsion between the steps, i.e. the step flow is expected to be stable. Actually, in the absence of both Ehrlich- Schwoebel effect and drift of adatoms there is no physical ground for step bunching instability. One should note, however, that this is true as long as the step velocity depends only on the width of the neighbouring terraces in the given moment, which means that the impact of the "history" of the terraces is neglected. To make clear this point some comments are necessary. For simplicity we consider the limit of fast surface diffusion and slow attachment-detachment at the steps. In this limit the adatom concentration is constant over the terrace. This concentration is low when the terrace is large because the atoms detached from the two steps spread over the large area of the terrace and, on the other hand, many atoms are lost by desorption from the large terrace. That is why "the history" of the terrace is essential. The adatom concentration depends not only on the width of the terrace in the moment but on its "past" as well. If the terrace was large in the "past" the adatom concentration is low. In contrast, the concentration is high if the terrace was small in the "past". This effect is completely neglected by the steady state treatment of the adatom concentration. On the other hand this effect provides a physical ground for oscillations of the terrace width. To verify this statement it is reasonable to go beyond the steady state treatment of the adatom concentration on the terraces.

Here, one should note the efforts of Ghez et al [1] and Keller et al [2] to clarify the stability of a step flow in the more complex situation when the surface diffusion of the adatoms is not assumed to be very fast. Their mathematical treatment is more sophisticated than the original Burton, Cabrera and Frank theory [3] in the sense that the quasi-static approximation is replaced by a Stefan-like problem for moving steps. This approach is relevant for the physical situation where the kinetics of crystallization (or evaporation of a crystal) is controlled by the relatively slow surface diffusion. The results obtained in [2] show a train of fast moving steps to be unstable. The treatment in [1, 2], however, neglects the step-step repulsion which is an important stabilizing factor.

Our aim here is to study the stability of the step flow in a presence of step-step repulsion. Since the case of fast surface diffusion and relatively slow attachment-detachment kinetics at the steps is easy to treat mathematically such a model provides a relatively simple way to study the linear stability of a step train. Last not least this model can be easily generalized to account for a drift of the adatoms as well as for the transparency of the steps. One should keep in mind, however, that the applicability of this model is limited to the regime characterized by fast surface diffusion and slow kinetics at the steps. The conclusions are definitely invalid in all those processes of sublimation and growth, which are controlled by slow surface diffusion.

The model of crystallization (evaporation) controlled by relatively slow attachment-detachment kinetics at the steps has an interesting non-linear dynamics which we study by numerical integration of the equations of step motion coupled with the equations describing the time evolution of the adatom concentration on the terraces. The step dynamics clearly manifests space and time periodicity (see Fig.1) appearing after some transition time.

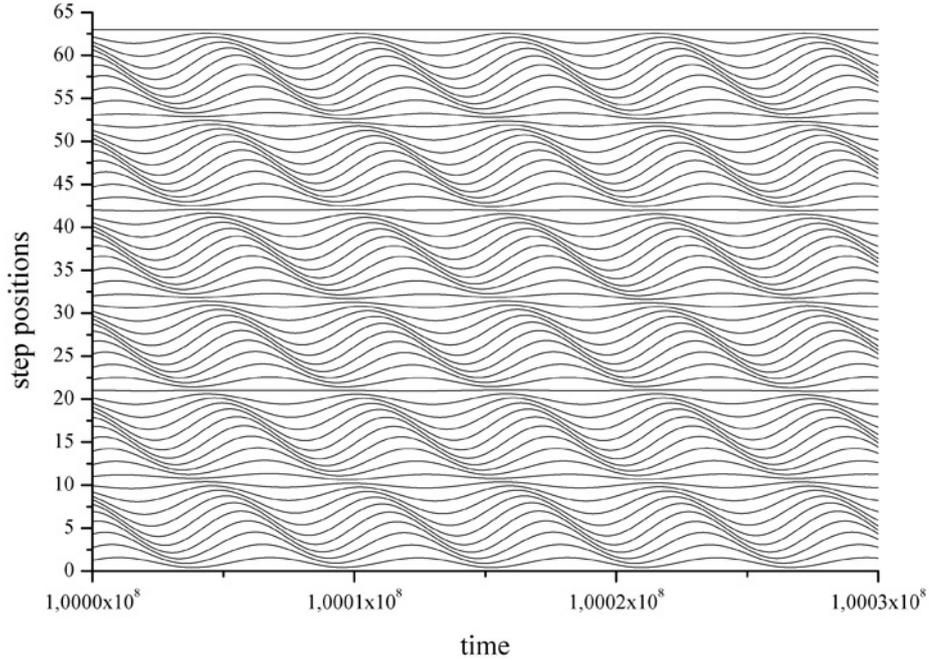

Fig. 1. Step trajectories in a frame moving with the first step. The space and time periodicity appears after some transition time.

2. Equations for terrace widths and adatom concentrations

Here we shall explain how the result shown in Figures 1 was obtained. To describe the processes at the crystal surface we introduce two types of variables – the concentration $n_i$ of adatoms on the $i-th$ terrace and the width $l_i = x_{i+1} - x_i$ of the $i-th$ terrace (as shown in Fig.2 $x_i$ denotes the position of the $i-th$ step). We assume the surface diffusion of adatoms to be very fast process in comparison with both the atom detachment from the steps and the adatom desorption from the crystal surface. In the limit of fast surface diffusion the concentration $n_i$ has a constant value over the whole terrace but it can vary with the sublimation time. It is relatively easy to write an expression for the time derivative of the adatom concentrations

$$\frac{dn_i}{dt} = -\frac{n_i}{\tau_s} - \frac{2K}{l_i} n_i + \frac{K}{l_i}\left[n_s^e(i+1) + n_s^e(i)\right] \quad (1)$$

where $\tau_s$ is the average life-time of an atom in a state of mobile adsorption on the crystal surface.

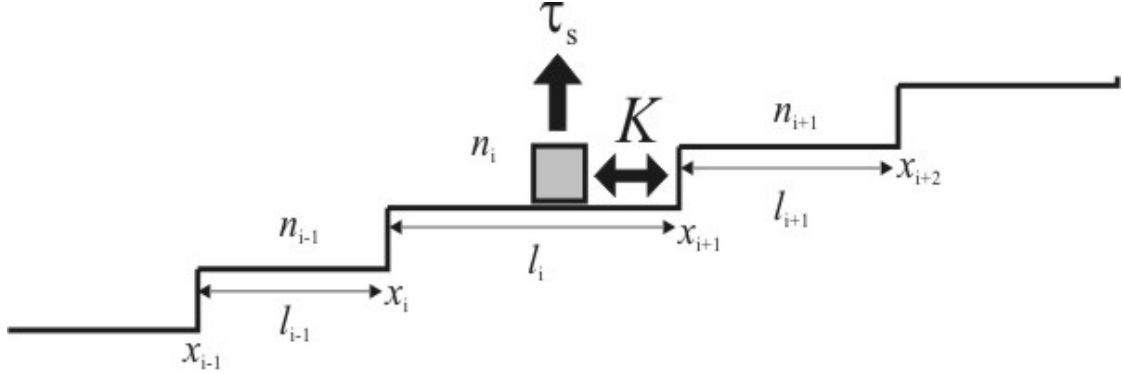

Fig. 2. Schematic view of vicinal surface with straight steps. The position of the $i-th$ step is denoted by $x_i$, the width of the $i-th$ terrace is $l_i = x_{i+1} - x_i$ and the concentration of adatoms on the $i-th$ terrace is $n_i$.

The exchange of atoms between the crystal phase and the dilute layer of adatoms is quantitatively characterized by the step kinetic coefficient K, defined by the expression $v_i = -K\Omega[n_i - n_s^e(i)] - K\Omega[n_{i-1} - n_s^e(i)]$. Here $v_i$ is the rate of the $i-th$ step and (according to Fig.2 the step rate $v_i$ is positive during sublimation and this explains the sign minus in the right hand side of the last expression). The rate of step motion is assumed to depend linearly on the local deviation $[n_i - n_s^e(i)]$ of the adatom concentration from its equilibrium value. As seen, the model we are going to treat here contains the same physics as the classical BCF model [3-5] with the additional simplifying assumption for a fast surface diffusion.

Going back to the equation (1) we see that it has a clear physical meaning – the adatom concentration decreases because of the desorption (the first term in the right hand side) and attachment of adatoms to the steps (the second term). The concentration $n_i$ of adatoms increases because of the detachment of atoms from the steps (the last two terms in the right hand side of Eq.(1)). We have used the notation $n_s^e(i)$ for the equilibrium concentration in the vicinity of the $i-th$ step to remind for the impact of the step-step repulsion on the local chemical potential and, therefore, on the equilibrium concentration of adatoms.

For a full description of the time evolution of the vicinal crystal surface we need a set of differential equations for the widths $l_i$ of the terraces. Having in mind that (see Fig.2)

$$\frac{dl_i}{dt} = \frac{dx_{i+1}}{dt} - \frac{dx_i}{dt} \quad (2)$$

and accounting for the linear dependence of the step velocity $\frac{dx_i}{dt}$ on the local undersaturation one can write

$$\frac{dl_i}{dt} = -K\Omega\{n_{i+1} - n_{i-1} - 2[n_s^e(i+1) - n_s^e(i)]\} \quad (3)$$

To obtain the final form of the set of differential equations (1) and (3) we should account for the contribution of the step-step interaction energy to the local chemical potential. The entropic and stress-mediated repulsion between the steps [6] provides for an interaction energy per unit length of the step described by

$$U = \frac{A}{l^2} \quad (4)$$

where $l$ is the interstep distance and A is estimated to be around $0.04(eVnm)$. Because of the above expression the equilibrium concentration of adatoms in the vicinity of the steps has the value $n_s^e$ only when the step distribution is perfectly regular ($l_i = l_{i-1} = l$). In the general case one has [7, 8]

$$n_s^e(i) = n_s^e\left[1 + \tilde{A}\left(\frac{1}{l_{i-1}^3} - \frac{1}{l_i^3}\right)\right] \quad (5)$$

where $\tilde{A} = \frac{2\Omega A}{kT}$.

Substituting eq.(5) into eqs.(1) and (3) we obtain

$$\frac{dn_i}{dt} = -\frac{n_i}{\tau_s} - \frac{2K}{l_i}n_i + \frac{2K}{l_i}n_s^e + \frac{K}{l_i}n_s^e \tilde{A}\left(\frac{1}{l_{i-1}^3} - \frac{1}{l_{i+1}^3}\right) \quad (6)$$

and

$$\frac{dl_i}{dt} = -K\Omega\left\{n_{i+1} - n_{i-1} + 2n_s^e \tilde{A}\left(\frac{1}{l_{i+1}^3} - \frac{2}{l_i^3} + \frac{1}{l_{i-1}^3}\right)\right\} \quad (7)$$

It is more convenient to use dimensionless variables $\tau = \frac{Kt}{l}$, $\eta_i = \frac{l_i}{l}$, $c_i = \frac{n_i}{n_s^e}$, $\bar{\eta} = \frac{\tilde{A}}{l^3}$. Thus the eqs. (6) and (7) turn into

$$\frac{dc_i}{d\tau} = -\frac{c_i}{\tau_s'} - \frac{2}{\eta_i}c_i + \frac{2}{\eta_i} + \frac{1}{\eta_i}\bar{\eta}\left(\frac{1}{\eta_{i-1}^3} - \frac{1}{\eta_{i+1}^3}\right) \quad (8)$$

and

$$\frac{d\eta_i}{d\tau} = -n_s^e \Omega \left\{ c_{i+1} - c_{i-1} + 2\overline{\eta} \left( \frac{1}{\eta_{i+1}^3} - \frac{2}{\eta_i^3} + \frac{1}{\eta_{i-1}^3} \right) \right\} \quad (9)$$

where $\tau_s^/ = \frac{\tau_s K}{l}$.

This set of 2N equations contains 3 physical parameters: $\tau_s^/$, $\overline{\eta}$ and the product $n_s^e \Omega$. It is instructive to briefly discuss the physical meaning of these parameters. To reveal the meaning of $\tau_s^/$ we consider a steady state concentration of adatoms on a vicinal surface with perfectly regular distribution of steps, i.e. $l_i = l_{i-1} = l$. For this case eq.(8) gives

$$c_0 = \frac{1}{1 + \frac{1}{2\tau_s^/}} \quad (10)$$

At $\tau_s^/ \gg 1$ one gets $c_0 \approx 1$ and therefore $n_i \approx n_s^e$, i.e. we have near to equilibrium conditions at the crystal surface. In the opposite limit $\tau_s^/ \ll 1$ we get $c_0 \approx 2\tau_s^/ \ll 1$ and therefore $n_i \ll n_s^e$. In this case we deal with sublimation under far from equilibrium conditions.

The physical meaning of the other two parameters is quit clear. The parameter $\overline{\eta}$ characterizes the strength of the step-step repulsion that is expected to have a strong impact on the step dynamics. Finally, the product $n_s^e \Omega$ gives the probability to find an atom at a given atomic site when the adatom concentration is equal to its equilibrium value. It should be pointed out that the values of $n_s^e \Omega$ can vary in a remarkably wide range. There are experimental evidences [9,10] that $n_s^e \Omega$ is of the order of 0.1 on the (111) surface of Si crystal at $900^0 C$ and $1250^0 C$. On the other hand for many metals the values of $n_s^e \Omega$ at temperatures close but below the melting point are expected to be in the interval $10^{-8} - 10^{-6}$.

3. Linear stability analysis

The simplest solution of the Eqs. (8) and (9) is $\eta_i = 1$ and $c_i = c_0 = \frac{1}{1 + 1/2\tau_s^/}$, which is an equidistant step distribution and the corresponding constant concentration of adatoms. The stability of this solution with respect to small fluctuations of the terrace size and adatom concentration is an interesting problem. To solve it we follow the routine procedure and consider small deviations from the equidistant step train and constant adatom concentrations, i.e., $\eta_i = 1 + \Delta\eta_i(\tau)$ and $c_i = c_0 + \Delta c_i(\tau)$. Substituting these expressions into Eqs. (8) and (9), making use of series expansion and keeping the linear terms only we get

$$\frac{d\Delta\eta_i}{d\tau} = -n_s^e \Omega \left[ \Delta c_{i+1} - \Delta c_{i-1} + 6\bar{\eta}\left(-\Delta\eta_{i+1} + 2\Delta\eta_i - \Delta\eta_{i-1}\right) \right] \quad (11)$$

$$\frac{d\Delta c_i}{d\tau} = -\frac{\Delta c_i}{\tau_s'} - 2\Delta c_i + 2(c_0 - 1)\Delta\eta_i + 3\bar{\eta}(\Delta\eta_{i+1} - \Delta\eta_{i-1}) \quad (12)$$

Following the routine we look for a solution of the type
$\Delta\eta_j = e^{ijq}\eta_q(\tau)$ and $\Delta c_j = e^{ijq+i\phi}c_q(\tau)$
where $q$ is a wave number and we already use $i$ to denote the imaginary unit and $j$ to denote the sequence number of the terrace. In addition we allow for a phase shift $\phi$ of the wave describing the adatom concentrations $\Delta c_j$ with respect to the wave describing the terrace widths $\Delta\eta_j$. In this way we arrive to a set of two differential equations

$$\frac{d\eta_q}{d\tau} = a_{11}\eta_q(\tau) + a_{12}c_q(\tau) \quad (13)$$

$$\frac{dc_q}{d\tau} = a_{21}\eta_q(\tau) + a_{22}c_q(\tau)$$

where the coefficients are given by the following expressions

$a_{11} = -12\bar{\eta}n_s^e\Omega(1-\cos q)$
$a_{12} = -2ie^{i\phi}n_s^e\Omega\sin q$
$a_{21} = 2e^{-i\phi}\left[-(1-c_0) + 3i\bar{\eta}\sin q\right] \quad (14)$
$a_{22} = -(2 + 1/\tau_s')$

This set of two linear differential equations has a solution of the type $e^{s\tau}$ where $s$ is a solution of
$(a_{11} - s)(a_{22} - s) - a_{12}a_{21} = 0 \quad (15)$
Therefore, the equidistant step distribution will be stable when the real part of $s$ is negative, but this regular distribution will be unstable if the real part of $s$ is positive.
In the Appendix we obtained the following approximate expression (see eq.(A11) for the real part of $s$

$$s_1 = \frac{6n_s^e\Omega\bar{\eta}q^2}{(2+1/\tau_s')\tau_s'}\left\{\left[\frac{8(n_s^e\Omega)/\tau_s'}{(2+1/\tau_s')^4 3\bar{\eta}} - 1\right] - q^2\left[\left[-\frac{1}{12} + \frac{1}{2}\tau_s'\right] + \frac{8}{9\bar{\eta}}\frac{(n_s^e\Omega)/\tau_s'}{(2+1/\tau_s')^4}\right]\right\} \quad (16)$$

It is instructive to look at this expression in the two limits: $\tau_s^/ \gg 1$ and $\tau_s^/ \ll 1$. In the first limit one gets

$$s_1 = \frac{n_s^e \Omega q^2 3\bar{\eta}}{\tau_s^/} \left\{ \left[ \frac{V}{V_{cr}} - 1 \right] - q^2 \left[ -\frac{1}{12} + \frac{1}{2}\tau_s^/ + \frac{(n_s^e\Omega)/\tau_s^/}{18\bar{\eta}} \right] \right\} \quad (17)$$

where $V = n_s^e \Omega l / \tau_s$ is the velocity of steps in a perfectly regular train and $V_{cr} = 6K\bar{\eta}$ is the critical velocity.

Equation (17) is an essential result. It shows that the step train is stable only when the velocity of steps is smaller than the critical value $V_{cr} = 6K\bar{\eta} = \frac{12K\Omega A}{kTl^3}$. Under this condition one has $s_1 < 0$ and, therefore, the amplitude of the fluctuations in the terrace width vanishes with the evaporation time. In contrast, the amplitude of the fluctuations increases when $V > V_{cr}$ because $s_1 > 0$ for small wave numbers and $\eta_q(\tau) \sim e^{s_1\tau}$ increases with the evaporation time. It is interesting to note that the critical velocity $V_{cr} = \frac{12K\Omega A}{kTl^3}$ is related to the initial velocity of the leading step of a bunch with interstep distance $l$ when the bunch is allowed to start relaxing. Really, since the leading step has a neighbour step only on the one side (the bunch is locked between two large step free surfaces) the equilibrium concentration in the vicinity of the leading step is $n_s^e(1) = n_s^e\left[1 + \frac{\tilde{A}}{l^3}\right]$. On the other hand the equilibrium concentration on the large step free crystal surface is $n_s^e$. Therefore a surface transport will start from the step (where the adatom concentration is higher) to the large step free surface and the step will move at a velocity

$$V_{step1} = K\Omega\left[n_s^e(1) - n_s^e\right] = \frac{K\Omega n_s^e \tilde{A}}{l^3} = \frac{2K\Omega n_s^e A\Omega}{l^3 kT} = \frac{1}{6}V_{cr}n_s^e\Omega \quad (18)$$

It is instructive to write the last expression in the form

$$V_{cr} = \frac{6V_{step1}}{n_s^e\Omega} \quad (19)$$

because we immediately see how the critical velocity depends on the model parameter $n_s^e\Omega$. As we already mentioned $n_s^e\Omega$ is of the order of 0.1 on the (111) surface of Si crystal at $900^0 C$ and $1250^0 C$ so that the critical velocity is 60 times larger than the initial velocity of the leading step during bunch relaxation. On the other hand for many metals the values of $n_s^e \Omega$ at temperatures below the melting point are in the interval $10^{-8} - 10^{-6}$ meaning that $V_{cr} \approx 10^7 V_{step1}$. In other words, the critical velocity for metals is very high.

Following the routine we find for the wave number of the most unstable mode

$$q_{max} = \left[ \frac{\frac{V}{V_{cr}} - 1}{2\left(\frac{\tau_s'}{2} + \frac{V}{3V_{cr}}\right)} \right]^{1/2} \quad (20)$$

Now we look at the other limit $\tau_s' \ll 1$. Following the same procedure we get

$$s_1 = 6 n_s^e \Omega q^2 \bar{\eta} \left\{ \left[ \frac{V}{V_{cr}} - 1 \right] - q^2 \left[ -\frac{1}{12} + \frac{1}{2}\tau_s' + \frac{V}{3V_{cr}} \right] \right\} \quad (21)$$

where

$V = K n_s^e \Omega$ and $V_{cr} = \frac{3\Omega A}{4kTK^2 \tau_s^3}$. The maximum of $s_1$ occurs at

$$q_{max} = \left[ \frac{\frac{V}{V_{cr}} - 1}{\frac{2V}{3V_{cr}}} \right]^{1/2} \quad (22)$$

Now we shall obtain an expression for the imaginary part of $s$ - the solution of the Eq.(15). This solution is given by Eq.(A1) and its imaginary part is

$\text{Imag} = \sqrt{r} \sin \frac{\Theta}{2}$

with

$r = \sqrt{(\text{Re})^2 + (\text{Im})^2}$, $\quad \Theta = \theta, \theta + 2\pi$, $\quad \cos\theta = \frac{\text{Re}}{r}$, $\quad \sin\theta = \frac{\text{Im}}{r}$.

Further we use the same approach like in the calculation of the real part $s_1$ to obtain

$$\text{Imag} = \frac{\text{Im}}{\sqrt{2\text{Re}}} \approx \frac{16 n_s^e \Omega (1 - c_0) \sin q}{(2 + 1/\tau_s')\sqrt{2}} \approx \frac{16 n_s^e \Omega q}{(2 + 1/\tau_s')^2 \sqrt{2\tau_s'}} \quad (23)$$

Now, the two limits: $\tau_s' \gg 1$ and $\tau_s' \ll 1$ of this expression are

$$\text{Imag} \approx \frac{2\sqrt{2} n_s^e \Omega q}{\tau_s'} \approx \frac{2\sqrt{2} n_s^e \Omega l q}{K \tau_s} = \frac{2\sqrt{2} V}{K} q \quad (24)$$

and

$$\text{Imag} \approx \frac{16 n_s^e \Omega \tau_s' q}{\sqrt{2}} = 8\sqrt{2} n_s^e \Omega K \tau_s q / l = \frac{8\sqrt{2} V \tau_s}{l} q \quad (25)$$

Let us note that the imaginary part of $s$ is associated with the propagation of compression waves which look as small bunches of steps.

4. Linear stability during growth

In a presence of non-zero deposition rate $R$ an additional term appears in the right hand side of eq. (8). Because of the dimensionless variables used in Eq.(8) this term reads $\frac{c_{st}}{\tau_s'}$ where $c_{st} = \frac{R \tau_s}{n_s^e}$. Obviously, when the deposition rate has its equilibrium value $R_e$ one obtains $c_{st} = \frac{R_e \tau_s}{n_s^e} = 1$. This means that $c_{st} < 1$ corresponds to undersaturation and $c_{st} > 1$ corresponds to supersaturation at the crystal surface. It is instructive to write down the equations for the dimensionless concentrations on the terraces in a presence of deposition rate

$$\frac{dc_i}{d\tau} = \frac{c_{st}}{\tau_s'} - \frac{c_i}{\tau_s'} - \frac{2}{\eta_i} c_i + \frac{2}{\eta_i} + \frac{1}{\eta_i} \bar{\eta} \left( \frac{1}{\eta_{i-1}^3} - \frac{1}{\eta_{i+1}^3} \right) \quad (26)$$

For the simplest case of equidistant step distribution $\eta_i = 1$ the steady state solution of the Eq. (26) is $c_0 = \frac{1 + c_{st}/2\tau_s'}{1 + 1/2\tau_s'}$. The small deviations $\Delta c_i(\tau)$ from this value ($c_i = c_0 + \Delta c_i(\tau)$) satisfy again the Eq.(12). Therefore, we can use the solution already obtained and simply replace $c_0$ with the new expression so that $1 - c_0 = \frac{1 - c_{st}}{1 + 2\tau_s}$. Substituting this expression into the second last equation of the Appendix we arrive to the conclusion that the step train is stable only when the velocity of the steps in the regular train $V$ is smaller than a critical value $V_{cr}$. In the limit $\tau_s' \gg 1$, i.e. under near to equilibrium conditions we obtain $V = n_s^e \Omega l |c_{st} - 1|/\tau_s$ and $V_{cr} = 6K\bar{\eta}/|c_{st} - 1|$.

It is reasonable to obtain an exact expression for the wave number of the most unstable mode without using a series expansion of $\sin q$ and $\cos q$ in eq.(A10). This wave number satisfies the equation

$$\frac{ds_1}{dq} \sim \left[ -1 + \frac{2\cos q}{(a_{11} - a_{22})} + \frac{4}{3} \frac{(n_s^e \Omega)(1-c_0)^2 2\cos q}{\bar{\eta}(a_{11} - a_{22})^3} \right] = 0 \quad (27)$$

Recalling that $1 - c_0 = \frac{1 - c_{st}}{1 + 2\tau_s'}$ and $a_{11} - a_{22} \approx (2 + 1/\tau_s')$ we restrict our considerations to the case $\tau_s' >> 1$ and obtain

$$\cos q_{max} = \frac{1}{1 + \frac{n_s^e \Omega (1 - c_{st})^2}{12 \bar{\eta} \tau_s'^2}} \quad (28)$$

In the limit $c_{st} \to 0$ one obtains the exact expression for the wave number of the most unstable mode in sublimation in vacuum.

5. Long time behaviour of vicinal surfaces

We study the non-linear dynamics of the steps at a vicinal surface by numerical integration of the equations (8) and (9) describing the time evolution of the terrace widths and the adatom concentrations. The set of equations (8) and (9) involves 2N variables because we use circling boundary conditions $\eta_{N+1} = \eta_1$ and $c_{N+1} = c_1$. Since only the terrace widths $\eta_i$ ($i=1, 2,...N$) are experimentally observable we shall plot the calculated data for the terrace widths only. To avoid overlapping of the plots $\eta_i(\tau)$ we define the quantities $\eta_i'(\tau) = (i-1)/2 + \eta_i(\tau)$, and plot them in Fig.3 for late times of intergration. These data for the terrace widths $\eta_i(\tau)$ are obtained by using the following values of the model parameters $\tau_s' = 50$, $\bar{\eta} = 10^{-5}$ and $n_s^e \Omega = 0.1$. Substituting these values into the expressions for $V$ and $V_{cr}$ results $\frac{V}{V_{cr}} = 33.3$, i.e. the step flow is highly unstable. As seen all terrace widths oscillate with the same frequency and amplitude. There is, however, some phase shift.

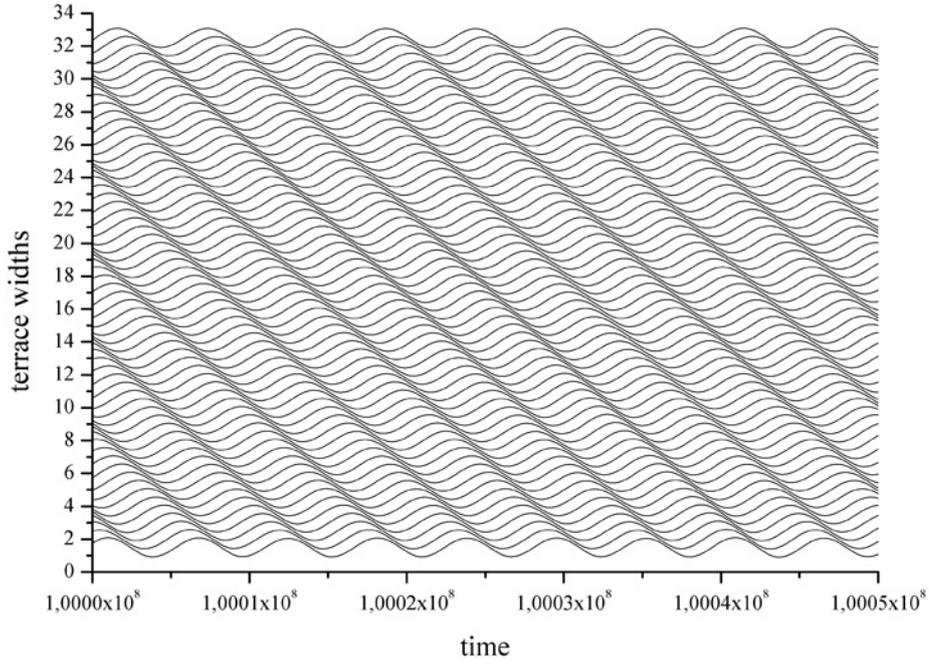

Fig. 3. Plot of $\eta_i'(\tau)$ at late times of integration. The model parameters are $\tau_s' = 50$, $\bar{\eta} = 10^{-5}$ and $n_s^e \Omega = 0.1$.

In fact, it is more instructive to plot the dimensionless trajectories ($\xi_i(\tau) = \frac{x_i(\tau)}{l}$) of the steps, defined in the following way $\xi_1(\tau) = 0$, $\xi_2(\tau) = \eta_1(\tau)$, $\xi_3(\tau) = \eta_1(\tau) + \eta_2(\tau)$,... $\xi_i(\tau) = \sum_1^{i-1} \eta_j(\tau)$. From the circling boundary conditions we have $\sum_1^N \eta_i(\tau) = N$ and, therefore, $\xi_{N+1}(\tau) = \sum_1^N \eta_i(\tau) = N$. It is clear now that the above presentation of the results from the numerical integration of equations (8) and (9) gives the step trajectories in a frame, moving with the first step. This presentation reveals very clearly the space periodicity in the step dynamics – if the n-th step repeats the motion of the first step the trajectory of the n-th step is a horizontal straight line (see Fig.4).

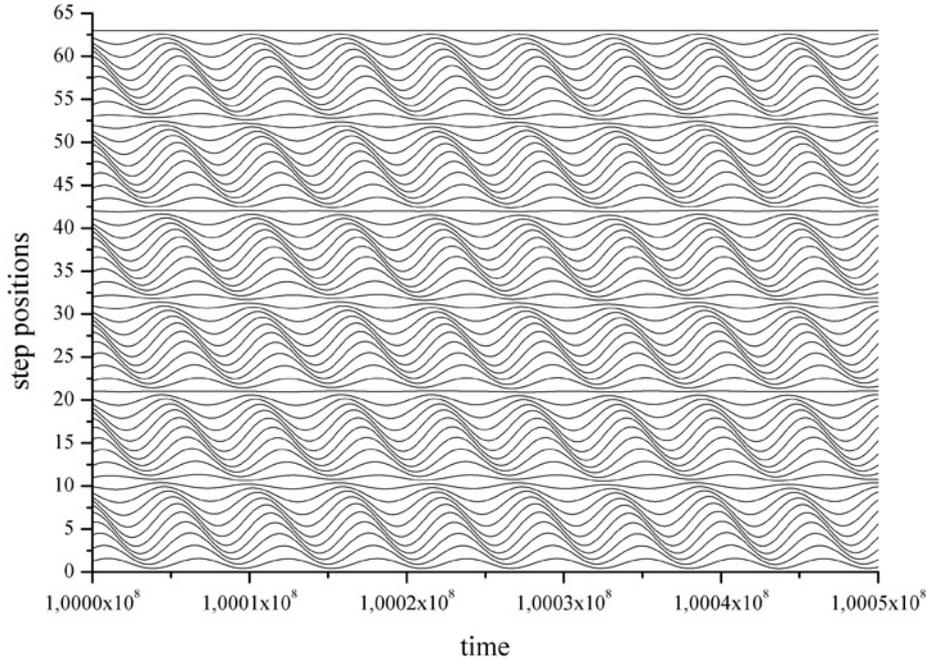

Fig. 4. Step trajectories in a frame moving with the first step. The model parameters are the same in Fig. 5.

Integrating the set of equations (8) and (9) and plotting the step trajectories in the way described above we obtained a variety of interesting patterns. It is interesting to compare the space periodicity manifested in Fig.5 with the predictions of the linear stability analysis. For the most unstable mode we use eq.(28) to obtain $q_{max} = 0.54$ and the corresponding wavelength $n_p = 11.6$ steps in good agreement with the numerical results shown in Figures 1 and 3.

The corresponding period $\tau_p$ of the most unstable mode is given by $\tau_p \mathrm{Imag} = 2\pi$ (according to eq.(24)), and substituting with $\tau_s^{/} = 50$, $n_s^e \Omega = 0.1$ and $q_{max} = 0.54$, for $\tau_p$ one gets $\tau_p \approx 2100$. This value is twice shorter than one we get from the numerical integration. The reason for this discrepancy is probably due to the fact that the numerical integration reproduce the step dynamics in the non-linear regime.

It is essential to note that the wavelength $n_p$ and the period of the compression waves depend on the initial configurations of the terraces at the vicinal surface.

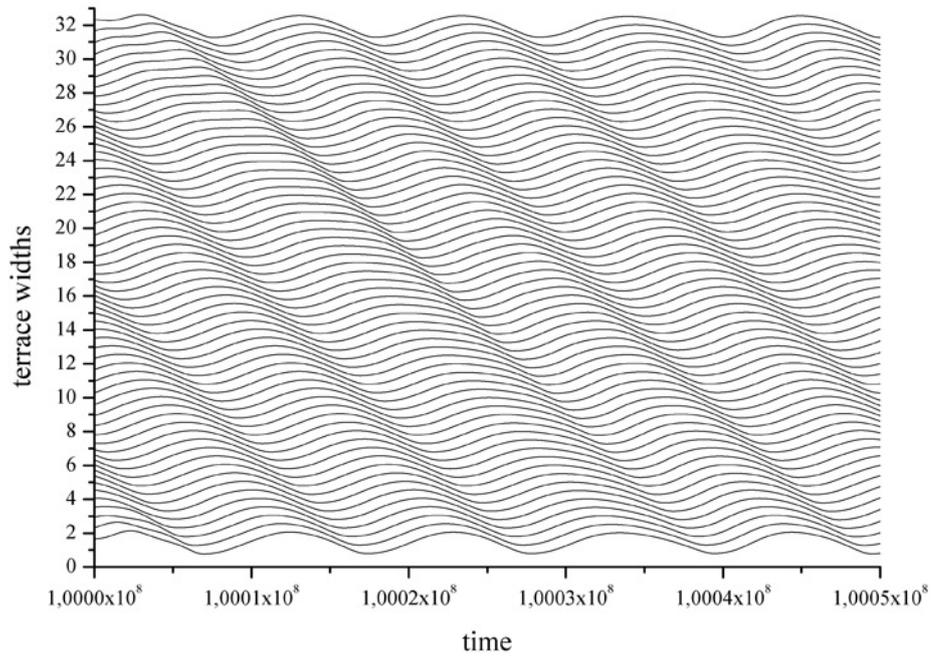

Fig. 5. Same as Fig.3, but integrated with very large initial fluctuations in the terrace widths.

For instance, Fig.5 shows terrace widths (the quantities $\eta'_i(\tau) = (i-1)/2 + \eta_i(\tau)$) for the same values of parameters as in Fig.3. The only difference is that we started the integration of the eqs.(8) and (9) from a configuration with very large fluctuations in the terrace widths.

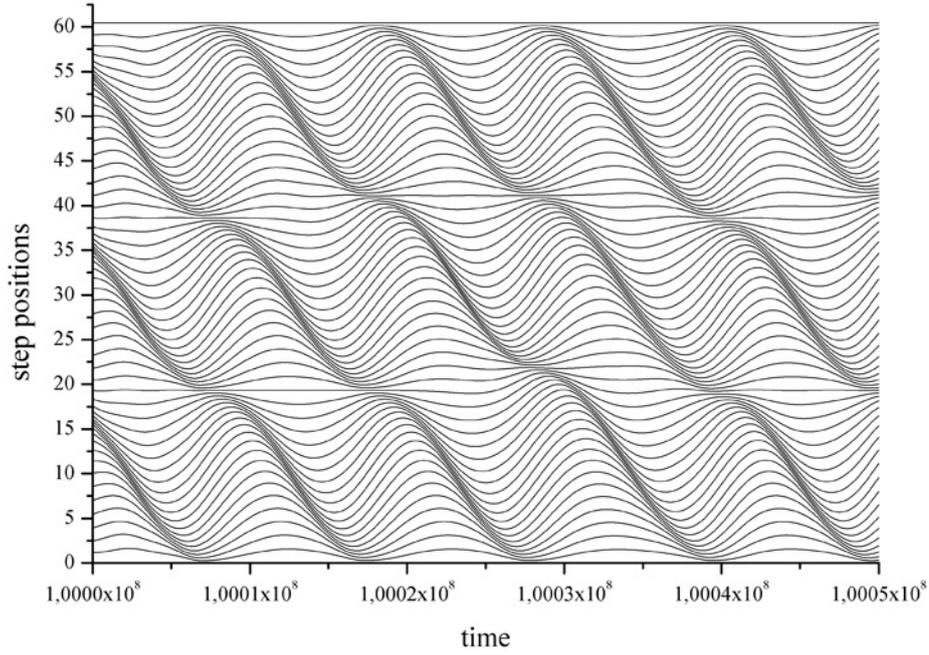

Fig. 6. Same as Fig.4, but integrated with very large initial fluctuations in the terrace widths.

The corresponding trajectories of the steps is shown on Fig.6 which presents very clear periodicity with a wavelength $n_p = 18$ steps which is different from the wavelength in Fig.4 although the values of the model parameters are the same as in Fig.1 and Fig.3.

6. Conclusion

We studied the step dynamics at vicinal surfaces during crystal sublimation and growth in absence of any destabilizing factors as Schwoebel barrier, electromigration of adatoms, etc. The central result of our work is that a critical velocity of the steps in the train exists which separates the stability and instability regimes. Instability occurs when the step velocity exceeds its critical value $V_{cr} = \dfrac{12K\Omega A}{kTl^3}$ . This instability differs from the usual step bunching where the average bunch size increases with the sublimation time. Integrating numerically the equations for the adatom concentrations and the motion of the steps we observed step density compression waves propagating at the vicinal surface of the crystal. It is interesting to note that the wave length as well as the frequency depend on the average size of the initial fluctuations in the terraces widths distribution. These waves are, in fact, small bunches of equal size. These small bunches do not coalesce and their size do not increase with the time of sublimation or growth.

The natural question is – what is the physical reason for this instability. We think the compression waves exist because the adatom concentration depends not only on the width of the terrace in the given moment but on the "past" of the terrace as well. If the terrace was large in the "past" the adatom concentration is low. In contrast, the concentration is high if the terrace was small in the "past". This effect can be accurately accounted for only by non-steady state treatment of the adatom concentration on the terraces. This is what we did in this paper for the limit of fast surface diffusion and slow kinetics of atom attachment-detachment at the steps.

7. Acknowledgement

This study is partially supported by Grant F-1413/2004 from the Bulgarian National Science Fund.

**APENDIX**

We start with the exact expression for the solution of Eq.(15)

$$s_{1,2} = \frac{1}{2}\left[(a_{11} + a_{22}) \pm \sqrt{(a_{11} + a_{22})^2 - 4(a_{11}a_{22} - a_{12}a_{21})}\right] \quad (A1)$$

The expression under the square root can be rewritten as $(a_{11} - a_{22})^2 + 4a_{12}a_{21}$. It has a real and imaginary part (we shall denote them by Re and Im). The real part of the square root itself we denote by Real and it is given by

$$\text{Real} = \sqrt{r}\cos\frac{\Theta}{2} \quad (A2)$$

where

$$r = \sqrt{(\text{Re})^2 + (\text{Im})^2}, \quad \Theta = \theta, \theta + 2\pi, \quad \cos\theta = \frac{\text{Re}}{r}, \quad \sin\theta = \frac{\text{Im}}{r} \quad (A3)$$

Substituting the two values of $\Theta$ into Eq.(17) we get

$$\text{Real}_1 = \sqrt{r}\cos\frac{\theta}{2}, \quad \text{Real}_2 = \sqrt{r}\cos\left(\frac{\theta}{2} + \pi\right) = -\text{Real}_1 \quad (A4)$$

It is, therefore, enough to find an expression for $\text{Real}_1$. It is convenient to rewrite the expression

$$\text{Real}_1 = \sqrt{r}\cos\frac{\theta}{2} = \sqrt{r\frac{1+\cos\theta}{2}} = \sqrt{\frac{r}{2} + \frac{\text{Re}}{2}} = \frac{1}{\sqrt{2}}\sqrt{r + \text{Re}} \quad (A5)$$

Further, having in mind that $\text{Im} = 4n_s^e\Omega(1-c_0)\sin q \ll \text{Re} \approx 4$ at small wave numbers, we write

$$\mathrm{Real}_1 = \frac{1}{\sqrt{2}}\left[\sqrt{(\mathrm{Re})^2 + (\mathrm{Im})^2} + \mathrm{Re}\right]^{1/2} = \frac{1}{\sqrt{2}}\left[\mathrm{Re}\sqrt{1 + \frac{(\mathrm{Im})^2}{(\mathrm{Re})^2}} + \mathrm{Re}\right]^{1/2} \quad (A6)$$

Making use of the series expansion of the square root (since $\frac{\mathrm{Im}}{\mathrm{Re}} \ll 1$) we get

$$\mathrm{Real}_1 \approx \frac{1}{\sqrt{2}}\left[\mathrm{Re}\left(1 + \frac{1}{2}(\mathrm{Im})^2/(\mathrm{Re})^2\right) + \mathrm{Re}\right]^{1/2} = \frac{1}{\sqrt{2}}\left[2\mathrm{Re} + \frac{1}{2}\frac{(\mathrm{Im})^2}{(\mathrm{Re})}\right]^{1/2} \quad (A7)$$

Let us now write the exact expressions for Re and Im

$$\mathrm{Re} = (a_{11} - a_{22})^2 + 48 n_s^e \Omega \bar{\eta} \sin^2 q \quad (A8)$$

$$\mathrm{Im} = 16 n_s^e \Omega (1 - c_0) \sin q \quad (A9)$$

Having in mind that $\bar{\eta} \leq 10^{-3}$, $\Omega n_s^e \leq 0.1$, $\sin q \ll 1$ for small wave numbers of interest, and for near to equilibrium conditions one has $(1 - c_0) \ll 1$ we conclude that the term $a_{22}$ is much larger than the other terms. This provides a ground to further simplify the expression for $\mathrm{Real}_1$

$$\mathrm{Real}_1 \approx \left[\mathrm{Re} + \frac{1}{4}\frac{(\mathrm{Im})^2}{(\mathrm{Re})}\right]^{1/2} \approx |a_{11} - a_{22}|\left\{1 + 24\frac{n_s^e \Omega \bar{\eta} \sin^2 q}{(a_{11} - a_{22})^2} + 32\frac{(n_s^e \Omega)^2 (1 - c_0)^2 \sin^2 q}{(a_{11} - a_{22})^4}\right\}$$

Now we go back to

$$s_1 = \frac{1}{2}[(a_{11} + a_{22}) + \mathrm{Real}_1] = \frac{1}{2}\left[2a_{11} + 24\frac{n_s^e \Omega \bar{\eta} \sin^2 q}{(a_{11} - a_{22})} + 32\frac{(n_s^e \Omega)^2 (1 - c_0)^2 \sin^2 q}{(a_{11} - a_{22})^3}\right]$$

and substituting $a_{11}$ with its expression (14) we arrive at

$$s_1 = 4 n_s^e \Omega \left[-3\bar{\eta}(1 - \cos q) + 3\frac{\bar{\eta} \sin^2 q}{(a_{11} - a_{22})} + 4\frac{(n_s^e \Omega)(1 - c_0)^2 \sin^2 q}{(a_{11} - a_{22})^3}\right] \quad (A10)$$

Restricting the treatment to small wave numbers we use series expansions $\sin q \approx q - \frac{1}{3!}q^3$ and $\cos q \approx 1 - \frac{1}{2!}q^2 + \frac{1}{4!}q^4$ and substitute them into the last expression. Thus we obtain

$$s_1 = 4 n_s^e \Omega \left[-3\bar{\eta}\left(\frac{1}{2}q^2 - \frac{1}{24}q^4\right) + 3\frac{\bar{\eta}\left(q^2 - \frac{1}{3}q^4\right)}{(a_{11} - a_{22})} + 4\frac{(n_s^e \Omega)(1 - c_0)^2\left(q^2 - \frac{1}{3}q^4\right)}{(a_{11} - a_{22})^3}\right]$$

The next approximation is $a_{11} - a_{22} \approx (2 + 1/\tau_s')$ because $\bar{\eta} \leq 10^{-3}$ and $\Omega n_s^e \leq 0.1$. Now the last expression takes the form

$$s_1 = \frac{4n_s^e \Omega q^2}{2 + 1/\tau_s'} \left\{ \left[ -3\bar{\eta}/2\tau_s' + 4\frac{(n_s^e \Omega)(1-c_0)^2}{(2+1/\tau_s')^2} \right] - q^2 \left[ \bar{\eta} \left[ -\frac{1}{8\tau_s'} + \frac{3}{4} \right] + \frac{4}{3} \frac{(n_s^e \Omega)(1-c_0)^2}{(2+1/\tau_s')^2} \right] \right\}$$

It is convenient to take the factor $3\bar{\eta}/2\tau_s'$ outside the brackets. Thus we obtain

$$s_1 = \frac{6n_s^e \Omega q^2 \bar{\eta}}{\left(2 + \frac{1}{\tau_s'}\right)\tau_s'} \left\{ \left[ 8\frac{(n_s^e \Omega)(1-c_0)^2 \tau_s'}{(2+1/\tau_s')^2 3\bar{\eta}} - 1 \right] - q^2 \left[ \left[ -\frac{1}{12} + \frac{\tau_s'}{2} \right] + \frac{8\tau_s'}{9\bar{\eta}} \frac{(n_s^e \Omega)(1-c_0)^2}{(2+1/\tau_s')^2} \right] \right\}$$

Since $c_0 = \dfrac{1}{1 + 1/2\tau_s'}$ and $1 - c_0 = \dfrac{1/2\tau_s'}{1+1/2\tau_s'} = \dfrac{1/\tau_s'}{2+1/\tau_s'}$ we substitute this into the expression for $s_1$ to finally obtain

$$s_1 = \frac{6n_s^e \Omega q^2 \bar{\eta}}{(2+1/\tau_s')\tau_s'} \left\{ \left[ \frac{8(n_s^e \Omega)/\tau_s'}{(2+1/\tau_s')^4 3\bar{\eta}} - 1 \right] - q^2 \left[ \left[ -\frac{1}{12} + \frac{\tau_s'}{2} \right] + \frac{8}{9\bar{\eta}} \frac{(n_s^e \Omega)/\tau_s'}{(2+1/\tau_s')^4} \right] \right\} \quad (A11)$$